\documentclass[aps,prc,twocolumn,superscriptaddress,floatfix]{revtex4-2}
\usepackage{graphicx}
\usepackage{amsmath}
\usepackage{amssymb}
\usepackage[colorlinks=true, breaklinks=true, linkcolor=blue, citecolor=blue, urlcolor=blue]{hyperref}

\begin{document}

\title{Differentiable emulation for high-dimensional Bayesian calibration of nuclear reaction models}

\author{Jin Lei}
\email{jinl@tongji.edu.cn}
\affiliation{School of Physics Science and Engineering, Tongji University, Shanghai 200092, China}
\affiliation{Southern Center for Nuclear-Science Theory (SCNT), Institute of Modern Physics, Chinese Academy of Sciences, Huizhou 516000, Guangdong Province, China}

\date{\today}

\begin{abstract}
Bayesian calibration of a coupled-channel reaction model requires many expensive solves, and the likelihood gradients needed by Hamiltonian Monte Carlo are absent from most production solvers. I present DREAM (Differentiable Reduced-basis Emulator with Automatic gradients for Markov-chain inference), a strategy that makes a reduced model differentiable without differentiating the original solver: parameter-dependent operators and solution snapshots are exported offline, compressed by singular value decomposition, and reconstructed with differentiable operations, so each likelihood evaluation returns the observables together with the exact gradient of the emulated likelihood. The construction applies to discretized linear models whose parameter dependence admits a low-rank, low-dimensional operator representation, and direct solver calculations verify the reduced model at the posterior points where it is used. Applied to continuum-discretized coupled-channels calculations of $d$+$^{58}$Ni elastic scattering at $20$~MeV, with eighteen central-potential parameters carried by twelve sampled coordinates, four chains yield 16\,000 retained samples in under an hour on one graphics processor, with an offline investment equivalent to about one thousand direct solves. Within the stated reaction model, continuum discretization, prior, and error assignment, the data determine the summed real and imaginary volume integrals as $772\pm13$ and $204\pm14$~MeV\,fm$^3$: the real integral lies $15\%$ below the summed Koning-Delaroche reference and is stable under a global factor-of-two rescaling of the assigned experimental errors, while the proton-neutron decomposition of each sum is left undetermined.
\end{abstract}

\maketitle

\section{Introduction}
\label{sec:intro}
Bayesian calibration asks how a data set, combined with a stated prior and a stated model, constrains the parameters of that model, and the answer is their joint posterior distribution. For nuclear reaction models the demand for such calibrations is growing, because uncertainty-quantified optical potentials are needed to propagate parametric uncertainty into the transfer, breakup, and capture calculations that connect rare-isotope experiments to astrophysical and applied quantities~\cite{Lovell2018,Pruitt2023,Beyer2026}. Gradient-free Markov chain Monte Carlo on fast forward models has produced such potentials in single-channel analyses~\cite{King2019PRL,Pruitt2023,Beyer2026}. What these analyses leave unused is the gradient of the likelihood. Gradient-guided samplers such as Hamiltonian Monte Carlo (HMC)~\cite{Duane1987,Hoffman2014,Neal2011} follow the local shape of the posterior instead of stepping at random, and their gain over random-walk methods grows as the parameters become more numerous and more correlated~\cite{Beskos2013,Huijser2022}. For a coupled-channel reaction model, where each likelihood evaluation is an expensive solve, these derivatives are not available: production codes do not provide them, and the emulators built from those codes have so far not supplied them either.

\begin{figure}[t]
\centering
\includegraphics[width=\columnwidth]{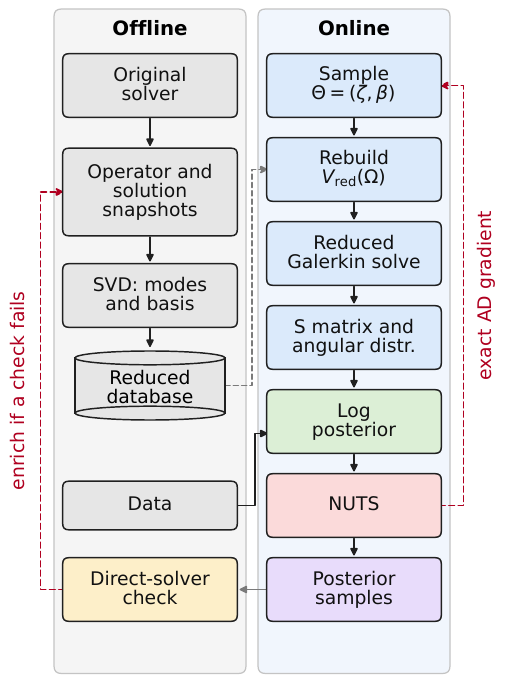}
\caption{Schematic of the DREAM construction. The sampled coordinates determine the optical-potential inputs $\boldsymbol{\Omega}$ of Eq.~(\ref{eq:pot}) and the fractional error scale $\beta$ of Eq.~(\ref{eq:like}). \emph{Offline:} the original solver exports parameter-dependent operators and solution snapshots, which are compressed. \emph{Online:} the emulator reconstructs $V_\mathrm{red}(\boldsymbol{\Omega})$, solves the Galerkin system, and returns the joint log-posterior gradient for NUTS. Direct calculations check posterior points before the reduced basis is accepted.}
\label{fig:framework}
\end{figure}

In this work I introduce DREAM (Differentiable Reduced-basis Emulator with Automatic gradients for Markov-chain inference), which supplies these derivatives without rewriting or automatically differentiating the production solver. The solver is used only to generate data offline; differentiation acts on the reduced model. The obstacle is that the potential geometry enters the operator nonlinearly: changing a radius or diffuseness changes every radial matrix element, rather than one coefficient of a fixed matrix. DREAM therefore has the solver export parameter-dependent matrices on a grid, together with solution snapshots that later form the projection basis; singular value decomposition (SVD) compresses these objects offline; and a reduced model built from differentiable operations reconstructs and solves them online, so automatic differentiation (AD) returns the posterior gradient with each evaluation (Fig.~\ref{fig:framework}). Although the demonstration below is a coupled-channels scattering calculation, the construction is stated for any discretized linear system $M(\boldsymbol{\Omega})\,\mathbf{c}=\mathbf{b}(\boldsymbol{\Omega})$ whose dependence on the calibrated parameters $\boldsymbol{\Omega}$ admits a low-rank, low-dimensional operator representation; Sec.~\ref{sec:discussion} states that condition precisely.

Related compression ideas underlie empirical interpolation~\cite{Barrault2004}, eigenvector continuation~\cite{Frame2018,Furnstahl2020}, and reduced-basis methods~\cite{Quarteroni2016,Drischler2023}, reviewed for quantum many-body and scattering problems in Ref.~\cite{Duguet2024}. Gradient-based Bayesian inference has been carried out for chiral nuclear interactions, whose low-energy constants multiply fixed operators: HMC posteriors have been obtained for the nucleon-nucleon sector up to fourth order~\cite{Svensson2022,Svensson2023} and, within an eigenvector-continuation emulator, with a correlated truncation-error model~\cite{Svensson2024}. In that affine setting the derivative of the emulated observable with respect to the couplings can even be assembled analytically, without automatic differentiation, from the same projected matrices that build the emulator~\cite{Melendez2021}. The present construction addresses the complementary case in which the parameters enter the operator nonlinearly through the potential geometry, so no fixed-matrix expansion exists in the physical parameters themselves; Sec.~\ref{sec:discussion} returns to the relation between the two routes. Emulators for optical and coupled-channel scattering with nonlinear parameter dependence have also been developed~\cite{Drischler2021,Odell2024,CatacorRios2025,LeiPINNECS,BiLNN}, but they have not been used to obtain a joint HMC posterior for a coupled-channel reaction observable. Speed has also been gained by changing the high-fidelity solver itself: wave-packet continuum discretization solves many scattering energies at once on a graphics processor at a quantified cost in method error~\cite{Miller2022}. When the parameters enter through a low-rank update of the Hamiltonian, few-body scattering and bound-state problems at fixed energy reduce exactly to small matrix problems, through a Woodbury identity or through a snapshot basis that spans the parameter-dependent solutions exactly~\cite{Heihoff2026}. DREAM combines operator compression, a differentiable Galerkin solve, posterior sampling, and direct verification against the original solver.

As a demonstration I apply DREAM to continuum-discretized coupled-channels (CDCC) calculations~\cite{Austern1987} of deuteron elastic scattering, in which the breakup continuum is explicitly coupled. CDCC combines expensive solves with nonlinear optical-potential geometry and strong parameter correlations, so it exercises every part of the construction. The reduced-operator decomposition and Galerkin solve are inherited from an earlier CDCC emulator~\cite{Lei2025}; here the online reconstruction and solve are differentiable, and the likelihood gradient is passed directly to the sampler.

I report a CDCC calibration of $d$+$^{58}$Ni elastic scattering at $E_\mathrm{lab}=20$~MeV in a reaction space with breakup partial waves $\ell \leq 4$ and continuum bins up to $20$~MeV, coupling up to 151 channels per total angular momentum. Four chains of the No-U-Turn Sampler (NUTS)~\cite{Hoffman2014}, the adaptive form of HMC used throughout, return $1.6\times10^4$ retained samples in under an hour on one data-center graphics processor. The sampled space has thirteen coordinates, high-dimensional relative to the two to six parameters usually varied in a single-system optical-model fit, though modest beside a global potential calibration~\cite{Pruitt2023}. The joint posterior singles out the total real and imaginary volume integrals as the combinations this observable constrains most tightly, and those are what I report, together with their sensitivity to the experimental-error assignment. At this dimension the measured efficiency gain of the gradient over a gradient-free sampler on the same emulator is modest; Sec.~\ref{sec:cost} reports it, and Sec.~\ref{sec:discussion} states where a larger gain is expected.

Sections~\ref{sec:bayes} to \ref{sec:online} set up the calibration problem, the CDCC forward model, and the offline and online stages of the construction. Sections~\ref{sec:setup} to \ref{sec:cost} give the data and prior, the verification against the original solver, the posterior results, and the cost of the whole campaign; Sec.~\ref{sec:discussion} discusses portability and limitations, and Sec.~\ref{sec:summary} summarizes.

\section{The calibration problem and the role of the gradient}
\label{sec:bayes}
The quantity of interest is the joint posterior density of the model parameters given the data, $p(\boldsymbol{\Theta}\,|\,\mathbf{y}) \propto \mathcal{L}(\mathbf{y}\,|\,\boldsymbol{\Theta})\,\pi(\boldsymbol{\Theta})$, where $\mathcal{L}$ is the likelihood of the measured angular distribution $\mathbf{y}$ and $\pi$ is the prior. Everything reported below is a property of this posterior, and therefore reflects the data, the prior, and the reaction model jointly; Sec.~\ref{sec:results} quantifies how the headline numbers respond when the error model inside $\mathcal{L}$ is changed.

A Gaussian likelihood compares the calculated Rutherford ratio $\mu_k$ at scattering angle $\theta_k$ with the measured value $y_k$:
\begin{align}
\ln\mathcal{L} = -\tfrac12 \sum_k \biggl\{ &\frac{[y_k - \mu_k(\boldsymbol{\Omega})]^2}{\sigma_k^2 + [\beta\,\mu_k(\boldsymbol{\Omega})]^2} \nonumber\\
&+ \ln\!\bigl[ \sigma_k^2 + [\beta\,\mu_k(\boldsymbol{\Omega})]^2 \bigr] \biggr\},
\label{eq:like}
\end{align}
where $\sigma_k$ is the uncertainty assigned to point $k$ and $\beta$ is an additional fractional error scale inferred jointly with the physics parameters~\cite{Kennedy2001,Beyer2026}. Both error contributions are treated as independent between angles; this diagonal error model does not represent correlated normalization, phase, or model-space errors. The logarithmic term is retained because the variance depends on the sampled $\beta$ and on $\mu_k(\boldsymbol{\Omega})$. The optical inputs obey $\boldsymbol{\Omega}=\boldsymbol{\Omega}(\boldsymbol{\zeta})$, where $\boldsymbol{\zeta}$ contains twelve sampled optical coordinates (Sec.~\ref{sec:setup} explains why twelve coordinates carry eighteen inputs), and the full sampled vector is $\boldsymbol{\Theta}=(\boldsymbol{\zeta},\beta)$.

HMC explores this posterior by integrating Hamiltonian trajectories in parameter space, and each leapfrog step requires the gradient of the log posterior with respect to every component of $\boldsymbol{\Theta}$. This is the step at which production reaction codes fail: they return the observables but not the derivatives, and finite-differencing them would cost one full solve per parameter per step, with a step-size noise floor set by the solver's own convergence tolerances. As the number of parameters grows and their correlations strengthen, the number of effectively independent samples obtained per likelihood evaluation by gradient-free methods collapses~\cite{Beskos2013}, and the stretch move of the affine-invariant ensemble sampler converges increasingly slowly as the dimension grows~\cite{Huijser2022}. HMC avoids that random-walk scaling and offers a practical route through this regime once an exact and cheap gradient is available; Sec.~\ref{sec:cost} measures what the route buys in the present calculation. Supplying that gradient for a coupled-channels reaction model is the purpose of the construction described in Secs.~\ref{sec:offline} and \ref{sec:online}.

\section{Forward model: CDCC for deuteron scattering}
\label{sec:cdcc}
The CDCC method~\cite{Austern1987} treats the deuteron as a $p+n$ pair whose relative motion is coupled to the target by the sum of a proton-target and a neutron-target interaction. The breakup continuum is discretized into momentum bins, and the total wave function is expanded over the bound state and the bin states $\phi_c$ of the pair, each coupled to the projectile-target orbital motion. Projecting the Schr\"odinger equation onto this basis gives the coupled radial equations
\begin{align}
\Bigl[-\frac{\hbar^2}{2\mu_R}\Bigl(\frac{d^2}{dR^2}-\frac{L_c(L_c+1)}{R^2}\Bigr) + \epsilon_c - E\Bigr] u_c(R) \nonumber\\
+ \sum_{c'} V_{cc'}(R)\, u_{c'}(R) = 0,
\label{eq:cdcc}
\end{align}
where $R$ is the projectile-target separation, $\mu_R$ the corresponding reduced mass, $E$ the total energy, and $c$ labels a channel: a bin (or the ground state) with internal energy $\epsilon_c$, internal orbital angular momentum $\ell_c$, and orbital angular momentum $L_c$, coupled to total angular momentum $J$. The coupling potentials $V_{cc'}(R)$ are matrix elements of the summed nucleon-target interactions between the internal states $\phi_c$ and $\phi_{c'}$, and they carry the entire dependence on the optical-potential parameters.

The fitted interaction is the central optical potential between each nucleon $x\in\{p,n\}$ of the deuteron and the target,
\begin{align}
U_x(r)={}&-V_v^x f(r;r_v^x,a_v^x)-iW_v^x f(r;r_w^x,a_w^x)\nonumber\\
&-iW_d^x g(r;r_d^x,a_d^x),
\label{eq:pot}
\end{align}
built from Woods-Saxon form factors
\begin{equation}
f(r;r_f,a_f)=\left[1+\exp\left(\frac{r-r_f A_T^{1/3}}{a_f}\right)\right]^{-1},
\label{eq:ws}
\end{equation}
with $A_T$ the target mass number and the surface shape $g(r;r_d^x,a_d^x)=-4a_d^x\,df(r;r_d^x,a_d^x)/dr=4f(1-f)$. Here $r_f$ is a reduced radius, so the physical radius is $r_f A_T^{1/3}$. The three depths, three reduced radii, and three diffusenesses for each nucleon give eighteen parameters, denoted collectively by $\boldsymbol{\Omega}$. The depths enter Eq.~(\ref{eq:cdcc}) linearly through $V_{cc'}$, but the geometry does not: changing a radius or diffuseness changes every element of the coupling matrix.

In the model space used throughout, the deuteron continuum has ten $2$-MeV bins up to $20$~MeV for each internal partial wave $\ell\leq4$, giving as many as 151 channels in a total-angular-momentum block, with $J\leq30$. The original solver expands each channel on a Lagrange-Legendre radial mesh of 180 points and imposes the scattering boundary conditions directly on the expansion coefficients~\cite{LeiDBMM2026}, so one total-$J$ block is a dense complex linear system of dimension up to $\sim\!2.7\times10^4$. A converged solve of all 31 blocks costs about half an hour on 16 processor cores. A calibration campaign of $10^5$ likelihood evaluations at this cost is out of reach, which is what motivates the reduced model; Sec.~\ref{sec:cost} returns to the full cost accounting.

\section{Offline stage: compression of the exported operator}
\label{sec:offline}
The present implementation requires an operator-export interface: the existing code must provide the parameter-dependent matrices, source vectors, solution snapshots, and boundary data needed by the reduced calculation. It does not require source-level AD: the export interface is the only code added, and no part of the internal solution flow is rewritten or differentiated.

Two compressions are applied, one to the operator and one to the solution space. Equation~(\ref{eq:pot}) holds separately for the proton and the neutron, so the interaction carries six form factors in all. They enter additively, and each is nonlinear in only its own geometry pair $(r_f,a_f)$. This structure replaces one eighteen-dimensional interpolation problem by six two-dimensional grids: the depths need no grid, and only the six geometry pairs are interpolated. The reduced optical interaction is
\begin{equation}
V_\mathrm{red}^J(\boldsymbol{\Omega}) = V_0^J + \sum_{f=1}^{6} \frac{\alpha_f}{\alpha_f^{(0)}} \,\Delta V_f^J(r_f, a_f),
\label{eq:decomp}
\end{equation}
where $V_0^J$ is the Coulomb interaction. It does not depend on $\boldsymbol{\Omega}$, but it is grouped with the optical terms because it is part of the interaction; the kinetic, centrifugal, and boundary terms are kept apart in the matrix $M_\mathrm{fixed}^J$ of the reduced system, Eq.~(\ref{eq:galerkin}) below. The six $\alpha_f$ are the depths $V_v^x$, $W_v^x$, and $W_d^x$ of Eq.~(\ref{eq:pot}), and each $\Delta V_f^J(r_f,a_f)$ is the coupled-channel matrix of form factor $f$ evaluated at a fixed reference depth $\alpha_f^{(0)}$, so that the prefactor $\alpha_f/\alpha_f^{(0)}$ rescales the matrix linearly to the sampled depth. The reference must be nonzero. For the four real-volume and surface form factors it is the corresponding Koning-Delaroche (KD02) depth~\cite{Koning2003}, evaluated for each nucleon at the corresponding nucleon energy; for the two volume-imaginary form factors, whose KD02 depths are small at this energy, the snapshot is taken at unit depth instead, which makes the prefactor a direct multiplication by the sampled depth. The two choices are algebraically equivalent whenever the reference is nonzero.

Each form factor is evaluated on a $20\times20$ grid covering the prior support in $(r_f,a_f)$, with 400 matrices per form factor. SVD of this snapshot ensemble gives
\begin{equation}
\Delta V_f^J(r_f, a_f) \approx \bar{V}_f^J + \sum_{m=1}^{K_f} \gamma_{f,m}(r_f, a_f) \, U_{f,m}^J,
\label{eq:pod}
\end{equation}
where $\bar{V}_f^J$ is the average of the 400 gridded matrices, subtracted before the decomposition so that the SVD acts on the fluctuations about it, $U_{f,m}^J$ are orthogonal matrix modes, and $\gamma_{f,m}$ are geometry-dependent scalar coefficients. A relative SVD tolerance of $10^{-4}$ retains $K_f=8$--$15$ matrix modes for each form factor. Only the scalars $\gamma_{f,m}$ are interpolated online. The source vectors are treated in the same way. This compression uses the additive form-factor structure of the interaction; a fully coupled parameter dependence would require a different coefficient model and cannot be reduced to six two-dimensional grids by smoothness alone.

The second compression reduces the linear system itself. SVD of coupled-channel solutions sampled over parameter space yields orthonormal solution modes for each total-$J$ block. Projecting Eq.~(\ref{eq:decomp}) onto $n_b$ of them reduces the system to dimension $n_b$, with $n_b=180$ here and the same rank used for every block. That this equals the radial mesh size is a coincidence; the two numbers count different things. The adequacy of the basis is established by comparison with the original solver, not by the rank alone.

The posterior also decides where that comparison matters. Testing a broad prior volume would consume the solves that the emulator is meant to avoid, whereas the posterior region is not known before inference. I therefore alternate inference and direct verification. If a posterior point fails, nearby full solutions are added and the basis is rebuilt, which moves the posterior and so calls for a fresh test. The loop stops when a basis passes at the posterior that basis itself produced, which took five rounds here; the accepted basis contains 728 solution snapshots. I measure the angular-distribution error in the same units used by the likelihood,
\begin{equation}
\chi^2_\mathrm{emu}=\sum_k\left[\frac{\mu_k^\mathrm{emu}-\mu_k^\mathrm{CDCC}}{\sigma_k}\right]^2.
\label{eq:emugate}
\end{equation}
The sum runs over the 64 data angles and is not divided by their number, and $\sigma_k$ is the assigned uncertainty alone, without the $\beta$ term of Eq.~(\ref{eq:like}), which can only enlarge the variance. An error of order unity can compete with the likelihood differences that separate parameter sets. I therefore require $\chi^2_\mathrm{emu}<1$ at representative points from the posterior produced by that basis. At this level the data cannot distinguish the emulated angular distribution from the direct one: if either were the truth, the other would lie within one unit of $\chi^2$ of it, which corresponds to a root-mean-square error per angle below $\sigma_k/8$. A sum can stay small while one angle is badly wrong, so I also require $\max_k|\mu_k^\mathrm{emu}-\mu_k^\mathrm{CDCC}|/\sigma_k<0.5$. Section~\ref{sec:verify} reports the outcome of this protocol for the accepted basis.

\section{Online stage: one gradient evaluation}
\label{sec:online}
Once the offline database is built, the original solver is not called during sampling. Dropping it is also what makes the gradient available. In the production code a change in $\boldsymbol{\Omega}$ propagates through every radial quadrature and every coupling matrix element, so the derivative can be reached only by carrying it through the entire solver. Equations~(\ref{eq:decomp}) and (\ref{eq:pod}) make the same dependence explicit instead, and the derivatives of the reduced operator can be read off directly,
\begin{align}
\frac{\partial V_\mathrm{red}^J}{\partial \alpha_f} &= \frac{\Delta V_f^J(r_f,a_f)}{\alpha_f^{(0)}},
\label{eq:dvdepth}\\
\frac{\partial V_\mathrm{red}^J}{\partial q_f} &= \frac{\alpha_f}{\alpha_f^{(0)}}\sum_{m=1}^{K_f} \frac{\partial \gamma_{f,m}}{\partial q_f}\,U_{f,m}^J ,
\label{eq:dvgeom}
\end{align}
with $q_f\in\{r_f,a_f\}$. The depth derivative is exact, and the geometry derivative needs only the derivative of a bilinear interpolant, available in closed form within each grid cell; the source vector $\mathbf{b}_\mathrm{red}^J$ is compressed the same way, so its derivatives follow from the same two equations. Everything downstream is elementary arithmetic on precomputed constants, so the chain rule supplies the composite derivative and AD evaluates it. Differentiability is built into the reduced representation rather than recovered from the solver, and it is exact only there: Eqs.~(\ref{eq:dvdepth}) and (\ref{eq:dvgeom}) differentiate the emulator that NUTS samples, not the original solver, and the difference between the two is what posterior verification measures. For each NUTS gradient evaluation, the online calculation performs three operations.

\emph{Step 1: Reconstruct the operator.} Bilinear interpolation gives $\gamma_{f,m}(r_f,a_f)$ at the proposed geometry, and Eq.~(\ref{eq:pod}) reconstructs $V_\mathrm{red}^J(\boldsymbol{\Omega})$. The interpolant is continuous and differentiable inside each grid cell; its derivative changes at cell boundaries. The AD gradient is therefore the exact derivative of this piecewise-bilinear emulator wherever that derivative exists. The emulator has three numerical approximations: projection onto the solution basis, SVD truncation of the form-factor matrices, and interpolation on the geometry grids.

\smallskip
\noindent\emph{Step 2: Solve the scattering equations.} With the reconstructed potential in hand, the online engine solves the reduced Galerkin system
\begin{equation}
\bigl[M_\mathrm{fixed}^J + V_\mathrm{red}^J(\boldsymbol{\Omega})\bigr]\,\mathbf{c}_J = \mathbf{b}_\mathrm{red}^J(\boldsymbol{\Omega})
\label{eq:galerkin}
\end{equation}
by direct linear algebra. Its solution $\mathbf{c}_J$ holds the coefficients of the scattering wave function in the $n_b$ solution modes, from which the elastic $S$-matrix elements are extracted analytically:
\begin{equation}
S_J(\boldsymbol{\Omega}) = 1 + \frac{2i}{H^+_J(\eta,kR)} \mathbf{u}_\mathrm{bnd}^\top \, \mathbf{c}_J(\boldsymbol{\Omega}),
\label{eq:smat}
\end{equation}
where $k$ is the elastic-channel wave number, $\eta$ the Sommerfeld parameter, $H^+_J(\eta,kR)$ the outgoing Coulomb-Hankel function at the matching radius $R$, and $\mathbf{u}_\mathrm{bnd}$ collects the elastic-channel values of the solution modes at that radius. The source vector in Eq.~(\ref{eq:galerkin}) carries the incident Coulomb wave, so $\mathbf{u}_\mathrm{bnd}^\top\mathbf{c}_J$ is the deviation of the elastic radial wave function from that incident wave at $R$, and Eq.~(\ref{eq:smat}) is the matching that returns $S_J$ relative to pure Coulomb scattering. The Coulomb phases are restored when the angular distribution is assembled from the $S_J$. Closed channels are matched to decaying Whittaker functions rather than outgoing waves. Projection onto an orthonormal solution basis also keeps the reduced systems well conditioned: across the KD02 center, the posterior mean, and eight posterior draws, the spectral condition number of the matrix in Eq.~(\ref{eq:galerkin}) ranges from $1.9\times10^3$ to $1.5\times10^4$ over all 31 blocks, with median $4.8\times10^3$, so the direct solve loses about four of the sixteen double-precision digits and the emulator error budget is set by the basis truncation and interpolation, not by conditioning.

\smallskip
\noindent\emph{Step 3: Evaluate the posterior and its gradient.} Reverse-mode AD evaluates the gradient of the log posterior with respect to $\boldsymbol{\Theta}=(\boldsymbol{\zeta},\beta)$ through the entire map, including the prior terms and the coordinate transformation $\boldsymbol{\Omega}(\boldsymbol{\zeta})$. Writing $A_J=M_\mathrm{fixed}^J+V_\mathrm{red}^J$, differentiation of Eq.~(\ref{eq:galerkin}) with respect to any component $\theta$ of $\boldsymbol{\Omega}$ gives
\begin{equation}
\frac{\partial \mathbf{c}_J}{\partial \theta} = A_J^{-1}\!\left[\frac{\partial \mathbf{b}_\mathrm{red}^J}{\partial \theta} - \frac{\partial A_J}{\partial \theta}\,\mathbf{c}_J\right],
\label{eq:implicit}
\end{equation}
with $\partial A_J/\partial\theta$ supplied by Eqs.~(\ref{eq:dvdepth}) and (\ref{eq:dvgeom}). The matrix to be inverted is the one already factored in the forward solve, and $S_J$ is linear in $\mathbf{c}_J$ by Eq.~(\ref{eq:smat}), so no new matrix has to be built or factored. Reverse mode reorganizes Eq.~(\ref{eq:implicit}) into one adjoint reduced-system solve per total-$J$ block at each gradient evaluation, using that same factorization. The number of adjoint solves does not grow with the number of fitted parameters; the assembly of the parameter derivatives still grows linearly.

With up to 151 channels, $n_b=180$ solution modes, and 31 total-$J$ blocks, the production timings below refer to one NVIDIA A800. A single-chain likelihood-plus-gradient evaluation is measured at $7.3$~ms in a tight loop, against $3.3$~ms for the forward observable alone, so the complete gradient with respect to all thirteen sampled coordinates adds about $1.2$ forward-evaluation equivalents to the forward solve, $2.2$ in total. Advancing the four chains together as one vectorized batch brings the cost to about $5$~ms per chain. At the posterior mean, on the interpolation grid used for the production run, the AD and central finite-difference derivatives of Eq.~(\ref{eq:like}) agree to a largest relative difference of $1.3\times10^{-6}$ and a mean of $1.6\times10^{-7}$ over all eighteen physical inputs, none of which has a vanishing derivative there. This checks the implementation at one point. The absence of divergences in all $1.6\times10^4$ retained transitions of the production run provides an empirical check that the piecewise-bilinear gradient remains numerically usable when trajectories cross interpolation-cell boundaries (Sec.~\ref{sec:verify}). The online calculation is implemented in the JAX framework with NUTS provided by NumPyro~\cite{JAX2018,NumPyro2019}; these tools are named here once for reproducibility, and nothing in the construction depends on them.

\section{Data, error model, and prior}
\label{sec:setup}
I use the $d$+$^{58}$Ni elastic angular distribution measured at $E_\mathrm{lab}=20$~MeV by Ermer et al.~\cite{dNi_data}, taken from the EXFOR compilation (accession F0709.024). The compilation quotes error bars at four of the sixty-four angles only. For the remaining points I assign absolute uncertainties of 0.005, 0.01, 0.02, or 0.05 in the Rutherford ratio: one common value across the forward region, and beyond it a value graded by the magnitude of the measured ratio, so that the assignment follows the diffraction pattern rather than the scattering angle. The result is an approximately constant relative uncertainty of median size $13\%$. The fractional term $\beta$ of Eq.~(\ref{eq:like}) has the same form, so it mainly tests whether that level is adequate; being non-negative it can enlarge an uncertainty that was set too small but never reduce one that was set too large. The intervals below are conditioned on this error model, and Sec.~\ref{sec:results} reports how they move when the assigned $\sigma_k$ are halved or doubled.

The spinless CDCC model treats the deuteron as a $p+n$ system bound by the central Gaussian interaction of Ref.~\cite{Austern1987} and the $^{58}$Ni target as inert. Intrinsic spins, spin-orbit interactions, and target excitation are omitted. The reduced Coulomb radius is fixed at 1.25~fm. These choices define the reaction model; the intervals below do not include uncertainty from the continuum discretization or the omitted degrees of freedom.

The prior center is the central part of KD02 for proton and neutron scattering from $^{58}$Ni at $E_N=E_d/2=10$~MeV~\cite{Koning2003}. The spin-orbit part of KD02 is omitted consistently with the spinless model. The eighteen physical inputs in Eq.~(\ref{eq:pot}) have truncated-normal marginal priors, with widths guided by the KDUQ nucleon optical-potential analysis~\cite{Pruitt2023}: $10\%$ for real depths, $25\%$ for surface depths, $3\%$ for reduced radii, and $5\%$ for diffusenesses. The KD02 volume-imaginary depths are $W_v^p=0.735$ and $W_v^n=0.790$~MeV at this energy, and I assign each a 2~MeV prior scale with the physical condition $W_v^{p,n}\geq0$. The real-volume depths are truncated at five prior standard deviations. Table~\ref{tab:priors} lists the parent mean, parent scale, and truncation interval of every marginal prior. The truncation intervals have three different origins: the real-volume depths carry the five-standard-deviation support, the imaginary depths are bounded below at zero by the physical requirement of absorption (a boundary the $W_v$ posterior does reach) and above by a value far outside the posterior, and the radii and diffusenesses are truncated at the spans of their interpolation grids, which is why the proton and neutron members of a geometry pair can carry different intervals where the two grids were built on different reference values.

\begin{table*}[t]
\centering
\caption{Truncated-normal priors, $\mathcal N(\mu,s)$ restricted to the stated interval. The left column group lists the proton-target parameters (superscript $p$) and the right group the neutron-target parameters (superscript $n$) of Eq.~(\ref{eq:pot}); each row pairs the proton and neutron members of the same term. Depths are in MeV; reduced radii and diffusenesses are in fm. For the surface-radius pair both truncation bounds lie beyond five parent standard deviations, so the shared-quantile map of Eq.~(\ref{eq:quantile}) returns $r_d^p$ and $r_d^n$ equal to within $3\times10^{-8}$~fm in every retained sample.}
\label{tab:priors}
\begin{ruledtabular}
\begin{tabular}{lccc@{\qquad}lccc}
Parameter & $\mu$ & $s$ & interval & Parameter & $\mu$ & $s$ & interval \\
\hline
$V_v^p$ & 55.302 & 5.530 & $[27.651,82.953]$ & $V_v^n$ & 49.535 & 4.953 & $[24.767,74.302]$ \\
$r_v^p$ & 1.1992 & 0.0360 & $[0.936,1.404]$ & $r_v^n$ & 1.1992 & 0.0360 & $[0.936,1.404]$ \\
$a_v^p$ & 0.6692 & 0.0335 & $[0.450,0.900]$ & $a_v^n$ & 0.6692 & 0.0335 & $[0.450,0.900]$ \\
$W_v^p$ & 0.735 & 2.000 & $[0,15]$ & $W_v^n$ & 0.790 & 2.000 & $[0,15]$ \\
$r_w^p$ & 1.1992 & 0.0360 & $[0.959,1.439]$ & $r_w^n$ & 1.1992 & 0.0360 & $[0.959,1.439]$ \\
$a_w^p$ & 0.6692 & 0.0335 & $[0.402,0.803]$ & $a_w^n$ & 0.6692 & 0.0335 & $[0.402,0.803]$ \\
$W_d^p$ & 7.904 & 1.976 & $[0,25]$ & $W_d^n$ & 7.500 & 1.875 & $[0,25]$ \\
$r_d^p$ & 1.2810 & 0.0384 & $[1.056,1.584]$ & $r_d^n$ & 1.2810 & 0.0384 & $[1.008,1.512]$ \\
$a_d^p$ & 0.5489 & 0.0274 & $[0.318,0.636]$ & $a_d^n$ & 0.5350 & 0.0267 & $[0.348,0.696]$ \\
\end{tabular}
\end{ruledtabular}
\end{table*}

The proton and neutron geometries are not sampled independently. KD02 assigns them the same central geometry for this target, apart from small prescribed offsets in the surface term, while a single deuteron angular distribution probes their folded sum. Giving the two geometries unrelated coordinates would let a proton-neutron shape compensation that this observable does not pin down enter the fit as if it were information. For each of the six proton-neutron geometry pairs $g$ I therefore sample one standard-normal coordinate $z_g$ and transform it separately for the proton and neutron parameters,
\begin{equation}
\Omega_i=F_i^{-1}[\Phi(z_g)], \qquad i\in\{p,n\},
\label{eq:quantile}
\end{equation}
where $F_i$ is the truncated-normal cumulative distribution of Table~\ref{tab:priors} and $\Phi$ is the standard-normal cumulative distribution. Members with the same parent mean and scale then track each other exactly, and unequal marginals, such as $a_d^p$ and $a_d^n$, retain their KD02 offset while moving through the same prior quantile. This preserves both KD02 marginal priors and their offsets, encodes the KD02 isoscalar geometry systematics, and removes relative proton-neutron shape changes that one deuteron angular distribution cannot determine. The six depths remain independent. The inference therefore has twelve optical coordinates controlling eighteen physical inputs, plus $\beta\sim\mathrm{HalfNormal}(0.05)$. The shared-quantile map is a strong prior modeling choice, unit correlation within each geometry pair, and the widths of the reported integrals are conditional on it. To measure that dependence I repeat the inference with all twelve geometry parameters sampled independently from their truncated-normal marginals, eighteen optical coordinates in place of twelve. The run converges (maximum split-$\hat R=1.014$, minimum effective sample size 299, no divergences) and gives $J_R/A_T=764\pm12$ and $J_W/A_T=205\pm11$~MeV\,fm$^3$ against $772\pm13$ and $204\pm14$ with shared coordinates (Sec.~\ref{sec:results}): the widths do not grow, the central values move by less than one combined standard deviation, and the extra freedom goes into the proton-neutron decomposition, whose anti-correlation strengthens from $-0.92$ to $-0.94$ for the real integral and from $-0.56$ to $-0.83$ for the imaginary one. The summed integrals are therefore not an artifact of the shared-quantile choice, whereas the individual proton and neutron contributions are sensitive to it, one more reason to report only the sums.

\section{Verification against the original solver}
\label{sec:verify}
The final run uses four distinct initial states from the posterior region supported by the direct-solver evaluations. Each chain readapts its step size and mass matrix for 2500 steps before retaining 4000 draws. The maximum split-$\hat R$ is 1.0051, the minimum effective sample size is 1358, no transition diverges, and the largest trajectory uses 95 leapfrog steps, well below the $2^{10}-1$ limit~\cite{Vehtari2021}.

Table~\ref{tab:gate} compares the reduced and direct CDCC angular distributions at thirteen points of the accepted posterior: the posterior mean; the retained sample closest to the pooled median in the whitened coordinates; the samples nearest the first and the ninety-ninth percentile of $V_v^p$, probing both tails of the least constrained depth; the retained sample with the largest error scale $\beta$; and two random draws from each of the four chains. Every point satisfies $\chi^2_\mathrm{emu}<0.3$ with a maximum pointwise error below $0.3\sigma_k$, well inside the acceptance criteria of Sec.~\ref{sec:offline}. These are spot checks spanning the bulk, both tails, and every chain of the posterior, not a certification of the full thirteen-dimensional posterior; they show that the emulator error is below the stated thresholds at representative points spanning the posterior bulk, its tails, and all four chains. The spread of the direct-CDCC $\chi^2$ in the last column simply reflects where the draws sit in the posterior; the largest value belongs to the largest-$\beta$ sample, whose inflated variance is what makes it retainable.

\begin{table}[b]
\centering
\caption{Direct-solver checks at posterior points. The second column is Eq.~(\ref{eq:emugate}), the third is the largest angular-distribution difference in units of the experimental error, and the last is the direct CDCC $\chi^2$ against the 64 data points. The first thirteen rows test the accepted posterior; the last row is the remote high-$V_v^p$ structure discussed in the text.}
\label{tab:gate}
\begin{ruledtabular}
\begin{tabular}{lccc}
Point & $\chi^2_\mathrm{emu}$ & $\max_k|\Delta|/\sigma_k$ & $\chi^2_\mathrm{CDCC}$ \\
\hline
Posterior mean & 0.154 & 0.159 & 74.37 \\
Posterior median-nearest & 0.186 & 0.201 & 70.73 \\
1\% lower tail of $V_v^p$ & 0.292 & 0.273 & 67.50 \\
99\% upper tail of $V_v^p$ & 0.109 & 0.093 & 101.71 \\
Largest-$\beta$ sample & 0.117 & 0.121 & 314.95 \\
Chain-1 draw a & 0.181 & 0.208 & 70.44 \\
Chain-1 draw b & 0.261 & 0.174 & 75.31 \\
Chain-2 draw a & 0.252 & 0.182 & 105.47 \\
Chain-2 draw b & 0.139 & 0.139 & 169.48 \\
Chain-3 draw a & 0.077 & 0.090 & 74.67 \\
Chain-3 draw b & 0.087 & 0.087 & 71.92 \\
Chain-4 draw a & 0.279 & 0.208 & 97.03 \\
Chain-4 draw b & 0.298 & 0.292 & 58.38 \\
High-$V_v^p$ median-nearest & 5.044 & 1.323 & 130.23 \\
\end{tabular}
\end{ruledtabular}
\end{table}

Exploration from the full prior found two remote structures in the reduced-model posterior, and both were adjudicated by the direct solver. A low-$V_v^p$ path favored by the reduced model is an extrapolation artifact of its own basis: direct CDCC lowers it by about 97 log-posterior units relative to the accepted region. The five-standard-deviation real-depth support of Table~\ref{tab:priors} excludes it; the accepted posterior has no samples within 1~MeV of that boundary, and its first-percentile point passes Table~\ref{tab:gate}. A high-$V_v^p$ structure near $69$~MeV is genuinely present in the reduced model, and no sampler crosses between it and the physical basin (Appendix~\ref{app:remote}). Direct CDCC does not support it: at its median-nearest point (the last row of Table~\ref{tab:gate}) the exact log posterior lies 16.578 units below that of the accepted posterior's median-nearest point (second row of Table~\ref{tab:gate}), and at twenty-four further draws spanning the structure, ten equilibrated samples from a full-prior chain confined there and fourteen from independently trapped ensemble walkers, it lies 15.5 to 32.5 units below that same point, with median 20.6. These deficits are pointwise density ratios of $e^{-15.5}$ to $e^{-32.5}$ at every direct evaluation made; turning them into an integrated posterior mass would require a solver-driven integration over the structure, which has not been done. The production posterior is therefore conditioned on the basin supported by the direct-solver evaluations, and the remote structure is characterized by direct calculation. The final four chains start from four different draws in the basin supported by the direct-solver evaluations and adapt their step sizes and dense mass matrices from scratch; no covariance from the exploratory runs is reused.

These checks have a definite scope. Posterior-point verification detects emulator errors in every region the reduced model reveals, whether physical or artificial, but it cannot rule out a region supported by the direct solver in which the reduced model is inaccurate enough that no chain ever concentrates there. No direct-solver scan of the prior volume was made. What anchors the surrogate across the prior is coverage rather than verification: the wide-prior and KD02-prior snapshots among the 728 that build the basis, and the check that the accepted basis reproduces the direct solution at the KD02 reference point itself ($\chi^2_\mathrm{emu}=0.035$, largest pointwise error $0.13\,\sigma_k$), a point far from the posterior. The reported posterior is therefore conditional on the basins explored and on that snapshot coverage.

\section{Results}
\label{sec:results}
\begin{figure}[t]
\centering
\includegraphics[width=\columnwidth]{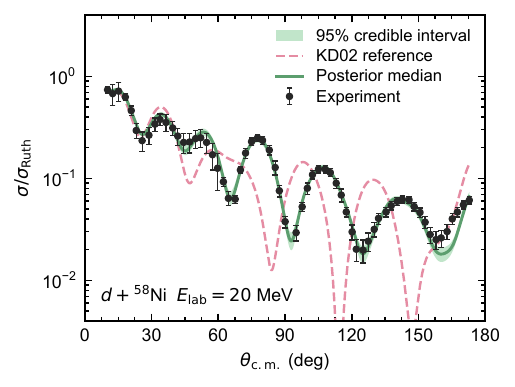}
\caption{Calculated $d$+$^{58}$Ni elastic angular distribution at $20$~MeV. Green band: 95\% credible interval from all 16\,000 retained posterior samples; solid green: posterior median; dashed rose: exact CDCC result at the KD02 central-potential reference for $E_N=10$~MeV; black points: experiment, with error bars showing the compilation-quoted uncertainties at the four angles that carry them and the assigned values of Sec.~\ref{sec:setup} elsewhere. The band propagates parameter uncertainty but does not add measurement noise or the fractional error term of Eq.~(\ref{eq:like}).}
\label{fig:predictive}
\end{figure}

Figure~\ref{fig:predictive} shows the posterior distribution of the calculated cross section. The 95\% band is computed from all 16\,000 retained samples; the same quantiles from a random 200-sample subset differ by at most $0.005$ in the Rutherford ratio, below the smallest assigned experimental uncertainty, so the band is insensitive to the number of draws. Direct CDCC at the posterior mean gives $\chi^2/N=1.16$ for the 64 points, evaluated with the assigned $\sigma_k$ alone and not with the $\beta$-inflated variance of Eq.~(\ref{eq:like}). The fractional error scale is $\beta=0.031\pm0.022$, with a 90\% interval of $[0.003,0.074]$. It remains compatible with a small additional independent error, but says nothing about correlated normalization, phase, or model-space errors. The uncalibrated KD02 central potentials give $\chi^2/N=51.8$ in the same CDCC model. This number does not test KD02 as a nucleon optical potential; it quantifies the failure of its unrenormalized proton-plus-neutron folding to describe this deuteron angular distribution and motivates the calibration.

I define the positive integral magnitudes per target nucleon by
\begin{align}
J_R^x/A_T &= \frac{4\pi}{A_T}\int_0^\infty V_v^x f_v^x(r)r^2\,dr,\\
J_W^x/A_T &= \frac{4\pi}{A_T}\int_0^\infty[W_v^x f_w^x(r)+W_d^x g_d^x(r)]r^2\,dr,
\end{align}
with $f_v^x(r)=f(r;r_v^x,a_v^x)$ and similarly for $f_w^x$ and $g_d^x$. Both are volume integrals in the $\int d^3r$ sense; $J_W^x$ collects the volume and surface absorption terms. I refer to their proton-plus-neutron sums as the total real and total imaginary integrals. Numerical summaries are posterior means and standard deviations unless an interval is stated explicitly.

\begin{figure}[!t]
\centering
\includegraphics[width=\columnwidth]{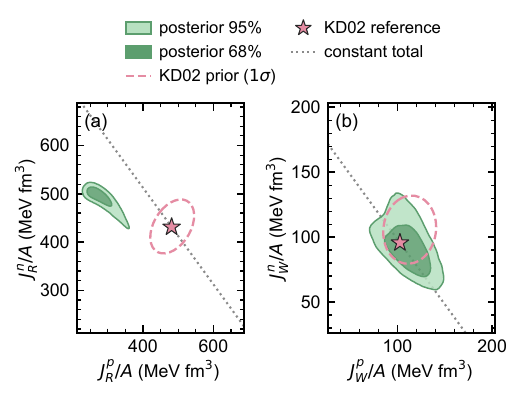}
\caption{Information in one elastic angular distribution: proton and neutron contributions to (a) the total real and (b) the total imaginary volume integral. The key identifies the posterior 68\% and 95\% regions, the one-standard-deviation covariance ellipse of the correlated KD02 prior, the KD02 reference, and the line of constant total through that reference.}
\label{fig:information}
\end{figure}

Figure~\ref{fig:information} summarizes the constrained combinations. The total real integral is $J_R/A_T=772\pm13$~MeV\,fm$^3$, and the total imaginary integral is $J_W/A_T=204\pm14$~MeV\,fm$^3$. The corresponding KD02 sums are 913 and 198~MeV\,fm$^3$. The real integral is therefore $15\%$ lower, whereas the imaginary integral differs by only $3\%$ and is statistically consistent with KD02. The real shift combines a smaller depth sum, $V_v^p+V_v^n=94.0\pm3.3$~MeV compared with the KD02 value 104.8~MeV, with a slightly smaller common radius, the direction of the classic depth-radius ambiguity of elastic scattering~\cite{Igo1958}. The imaginary depth sums are $W_d^p+W_d^n=18.2\pm1.9$~MeV and $W_v^p+W_v^n=1.73\pm0.83$~MeV. The volume-imaginary depths are small and their posterior reaches the physical wall at $W_v=0$, with $6\%$ and $9\%$ of the proton and neutron samples falling within $0.1$~MeV of it, so the two imaginary terms are not separately identified.

Panels (a) and (b) of Fig.~\ref{fig:information} also show how the two contributions are shared. They are strongly anti-correlated, with coefficients $-0.92$ for the real integral and $-0.56$ for the imaginary one, so the posterior follows ridges of nearly constant total strength. I therefore report the two sums rather than the four separate contributions. Their separation depends on the reaction model and is not pursued here. The posterior of the twelve sampled optical coordinates is shown in Appendix~\ref{app:corner}; the summed volume integrals above and their anti-correlated decompositions are the combinations that carry the physics conclusions.

Because the assigned $\sigma_k$ are an author-side reconstruction rather than a published covariance, I repeat the entire inference with all assigned errors doubled and with all assigned errors halved, leaving everything else unchanged. With doubled errors the run converges as cleanly as the production run (maximum split-$\hat R=1.0004$, no divergences) and returns $J_R/A_T=777\pm9$ and $J_W/A_T=215\pm10$~MeV\,fm$^3$: the real integral moves by $0.7\%$ and the imaginary one by $5\%$, each within one combined standard deviation of the production values, and the $15\%$ real-integral deficit relative to KD02 is unchanged; the error scale is $\beta=0.029\pm0.022$, consistent with the production value. The narrower widths under doubled errors do not indicate more information from the data. With the weaker likelihood the posterior slides along the proton-neutron degeneracy ridge toward the prior center for the split, from $V_v^p=34.2\pm2.5$ and $V_v^n=59.8\pm4.6$~MeV to $44.0\pm4.5$ and $46.2\pm4.5$~MeV; each real-integral contribution broadens, from $282\pm29$ and $490\pm20$ to $379\pm38$ and $398\pm37$~MeV\,fm$^3$, while their anti-correlation tightens from $-0.92$ to $-0.97$, and their strongly anti-correlated sum narrows as a result. Both posteriors remain far inside the prior, whose summed real integral is $915\pm99$~MeV\,fm$^3$. With halved errors the run converges equally cleanly (maximum split-$\hat R=1.005$, minimum effective sample size 1121, no divergences) and returns $J_R/A_T=771\pm18$ and $J_W/A_T=208\pm22$~MeV\,fm$^3$, while the sampled error scale rises to $\beta=0.122\pm0.022$, which restores the total effective uncertainty of Eq.~(\ref{eq:like}) to nearly its production level: the likelihood itself rejects an error assignment tighter than the scatter of the data supports. The headline integrals are therefore stable against a global factor-of-two rescaling of the assigned experimental errors in either direction. This probes only the global scale: angle-dependent assignments, normalization offsets, and correlated covariances are outside the tested family, and the diagonal error model of Sec.~\ref{sec:bayes} remains a stated condition of the analysis.

\section{Computational cost and comparison with gradient-free sampling}
\label{sec:cost}
The cost of the calibration separates into an offline investment and an online sampling cost, and the two should be reported together. A converged direct CDCC solve of the full model space costs about 29 minutes on 16 processor cores, roughly $7.7$ core-hours per likelihood evaluation. A gradient-free campaign of $10^5$ such evaluations would therefore cost of order $10^6$ core-hours, which is what makes direct sampling impractical and motivates the emulator.

The offline stage consists of three parts. First, the 728 solution snapshots that build the reduced basis: each snapshot is one direct solve, accumulated over the five enrichment rounds of Sec.~\ref{sec:offline}, for roughly $5\times10^3$ core-hours in total. Second, the six form-factor grids: $6\times400$ reduced-operator exports, which reuse the already-factored reduced basis and cost a few core-minutes per grid point, roughly $3\times10^2$ core-hours. Third, the direct verification solves of Sec.~\ref{sec:verify}, a few tens of full solves across the enrichment history, roughly $3\times10^2$ core-hours. The complete offline database therefore costs the equivalent of roughly $10^3$ direct CDCC evaluations, about two orders of magnitude less than a single gradient-free campaign on the original solver, and it is a reusable asset: the same database serves every subsequent inference at this energy, including the two error-model reruns of Sec.~\ref{sec:results}, which required no new offline work.

The online cost is then almost negligible. One likelihood-plus-gradient evaluation costs $7.3$~ms on one NVIDIA A800 ($3.3$~ms for the observable alone), and the four-chain production run of $1.6\times10^4$ retained samples completed in $3.1\times10^3$~s of wall time, dominated by the leapfrog integrations of NUTS. The doubled-error rerun, whose posterior is closer to Gaussian, completed in $1.8\times10^3$~s. For comparison, the non-differentiable Fortran reduced-basis emulator of Ref.~\cite{Lei2025} evaluates the same observable in about ten seconds per call on a multicore node; at that rate the leapfrog trajectories of this run alone would cost weeks, before accounting for the absent gradient. The wall-clock figures on different hardware are not directly commensurable; the figure that transfers between machines is the evaluation count. Including warmup, the production run used about $6\times10^5$ gradient evaluations, each costing $2.2$ forward evaluations, for a minimum effective sample size of 1358, and every one of those evaluations would otherwise have been a direct solve.

\begin{figure}[t]
\centering
\includegraphics[width=\columnwidth]{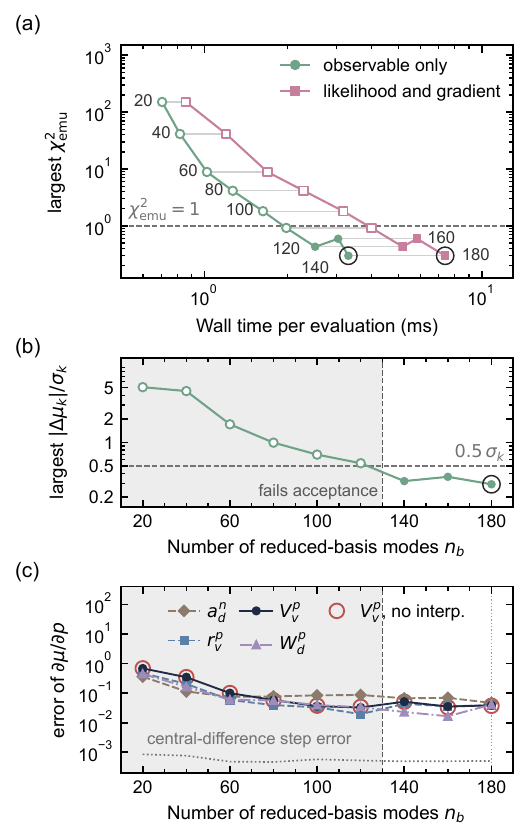}
\caption{Accuracy versus cost as the number $n_b$ of reduced-basis modes varies; all ranks are truncations of the production basis. (a) Largest $\chi^2_\mathrm{emu}$ over the thirteen posterior points of Table~\ref{tab:gate} against wall time per evaluation on one A800, for the observable (circles) and for a simplified likelihood with its gradient (squares); numbers give $n_b$, gray lines join the two points of one rank, open symbols fail an acceptance criterion, and circles mark the production rank. (b) Largest pointwise error $|\mu_k^\mathrm{emu}-\mu_k^\mathrm{CDCC}|/\sigma_k$; gray shading marks the ranks below the first one that passes both criteria. (c) Relative error, in a $1/\sigma_k$-weighted norm over the 64 angles, of the emulator derivative of $\mu_k$ at the posterior mean against central differences of direct CDCC calculations; red rings (``no interp.'') repeat the $V_v^p$ test with the form-factor operators computed directly at that geometry instead of interpolated, and the dotted curve is the central-difference step error measured on the emulator.}
\label{fig:cat}
\end{figure}

Figure~\ref{fig:cat} shows how accuracy and cost trade against each other as the rank of the solution basis changes, with panel (a) in the computational-accuracy-versus-time (CAT) form used for scattering emulators~\cite{Odell2024,CatacorRios2025}. Because the basis columns are ordered by singular value and every reduced quantity is a projection onto them, the emulator with $n_b<180$ modes is obtained exactly by keeping the leading $n_b$ rows and columns of each reduced matrix and vector, with no new offline work; at $n_b=180$ it reproduces the thirteen posterior rows of Table~\ref{tab:gate}. Over these points the largest $\chi^2_\mathrm{emu}$ falls from 151 at $n_b=20$ to 0.30 at $n_b=180$, while the time per evaluation grows by a factor of 4.7 for the observable and 8.6 for a likelihood with its gradient. Both acceptance criteria are first met at $n_b=140$ ($\chi^2_\mathrm{emu}\leq0.43$, largest pointwise error $0.32\,\sigma_k$); at $n_b=120$ the summed error is already below one but one angle exceeds $0.5\,\sigma_k$, the case the pointwise criterion is there to catch. The production basis carries a margin of about 1.4 in $\chi^2_\mathrm{emu}$ at 1.3 times the forward cost of the smallest passing one. The ranks below it are not usable, but they show two features of the problem: the error decreases slowly, on average by a factor of 1.8 per twenty modes between $n_b=60$ and 180, so the coupled-channel solutions in the posterior region need more than a hundred modes; and the online cost grows only by a factor of 3.2 for the observable and 4.4 with the gradient while $n_b$ triples, far less than the cubic scaling of a dense solve. Accuracy is cheap to buy online; the expensive part is the offline snapshot set. Panel (c) compares the emulator derivatives with the original solver: central differences of direct CDCC calculations at the posterior mean, with steps of 0.05 prior standard deviations in $V_v^p$, $r_v^p$, $W_d^p$, and $a_d^n$, agree with the automatic-differentiation derivatives of the emulator to 3.9 to 4.7\% at the production rank, in a $1/\sigma_k$-weighted norm over the 64 angles. The agreement levels off at a few percent from $n_b\approx60$ onward, far above the step error of the reference itself: halving the step changes the direct-solver derivatives by at most $4\times10^{-4}$. The sampling is not affected, since NUTS uses the exact gradient of the emulated posterior. Computing the form-factor operators directly at the posterior-mean geometry, instead of interpolating them from the geometry grid (red rings in Fig.~\ref{fig:cat}c), lowers $\chi^2_\mathrm{emu}$ at that point from 0.154 to 0.049 but leaves the derivative discrepancies for $V_v^p$, $r_v^p$, and $W_d^p$ at 2.7 to 3.7\% (only $a_d^n$ improves, from 4.7 to 1.7\%), so the plateau comes from the solution space rather than the interpolation: the snapshots span the solutions, whereas the derivative of the full-space solution $\boldsymbol{\psi}$ of $M\boldsymbol{\psi}=\mathbf{b}$, $\partial\boldsymbol{\psi}/\partial\theta=M^{-1}[\partial\mathbf{b}/\partial\theta-(\partial M/\partial\theta)\,\boldsymbol{\psi}]$, also has components outside that span. Adding sensitivity snapshots to the basis is the natural extension. The timings were measured on one A800, and at $n_b=180$ the forward time reproduces the 3.3~ms of Sec.~\ref{sec:online}; the gradient timings use a simplified objective, a fixed-error Gaussian likelihood differentiated with respect to the eighteen physical inputs, and serve as a relative benchmark between ranks.

To make the comparison with gradient-free sampling concrete, I ran an affine-invariant ensemble sampler~\cite{ForemanMackey2013} on the identical thirteen-coordinate posterior, with the same reduced model and the same processor: 64 walkers for $10^5$ steps, $6.4\times10^6$ likelihood evaluations, ten times the model calls of the production NUTS run. The ensemble is cold-started from the prior, so its proper gradient-based comparator is the full-prior NUTS run of Appendix~\ref{app:remote}, which received the same starting information and a similar budget; the production run starts inside the basin supported by the direct-solver evaluations, and the efficiency figures below are conditional on that region.

Neither sampler class crosses the barrier around the remote high-$V_v^p$ structure. Eighteen of the 64 walkers start in its basin of attraction and never leave it, so the pooled ensemble weights the two structures by the initial walker allocation, and the pooled trace diagnostics do not distinguish this from convergence, although per-walker inspection does (Appendix~\ref{app:remote}). The full-prior NUTS run populates both structures as well and cannot mix between them; its between-chain $\hat R=7.35$ flags the disconnection only because four independently initialized chains are compared, a protocol equally necessary for ensemble runs. On a multimodal surrogate posterior, neither sampler's internal diagnostics replace overdispersed multi-start initialization plus direct-solver adjudication of every structure found, which is the protocol of Sec.~\ref{sec:verify}. Within the physical basin the two samplers agree: the 46 untrapped walkers reproduce the converged NUTS marginal to one percent ($V_v^p=34.2\pm2.4$ versus $34.2\pm2.5$~MeV), an independent cross-check of the production posterior, and the efficiency comparison is close, NUTS ahead by about 2.4 in effective samples per model call and near parity once each gradient evaluation is billed at its 2.2-forward-evaluation cost. The gradient improves efficiency per model call and is exact for the emulated posterior; it does not by itself solve multimodality.

\section{Discussion}
\label{sec:discussion}
The main methodological result is a differentiable reduced operator constructed from matrices exported by an existing reaction solver. It supplies the exact derivative of the emulator posterior for one extra back-substitution per total-$J$ block, reusing the factorization from the forward solve, and posterior checks against the original solver determine whether the solution basis must be enriched. The CDCC calculation demonstrates the procedure for twelve correlated optical coordinates, one error coordinate, and eighteen physical potential inputs.

The construction should be placed against the existing gradient routes. For chiral interactions the couplings multiply fixed operators, so the emulator matrices are polynomials in the parameters and their derivatives are available analytically, without AD~\cite{Melendez2021}; HMC calibrations of the low-energy constants exploit exactly this structure~\cite{Svensson2022,Svensson2023,Svensson2024}. When the affine update is also of low rank, the fixed-energy problem reduces exactly, through a Woodbury identity to a matrix equation whose size is that rank, or to a snapshot basis at most one vector larger, and the emulator adds no error beyond that of the discretized problem even far from its snapshots~\cite{Heihoff2026}. A local potential is diagonal in the radial mesh index, and a change of any of its parameters, the depth included, alters it at essentially every mesh point, so the rank of the update grows with the mesh size and no small exact reduction of this kind exists. The present emulator therefore carries its own projection, truncation, and interpolation errors, which must be checked against the original solver (Sec.~\ref{sec:verify}). Equations~(\ref{eq:dvdepth}) and (\ref{eq:dvgeom}) show that the present construction recovers the same kind of closed-form operator derivatives for the non-affine case, at the price of the offline geometry grids; AD then automates the remaining chain rule from the operator to an arbitrary downstream observable, so no new derivation is needed when the observable changes. The two routes are therefore complementary rather than competing: where an affine expansion exists it should be used, and where it does not, the exported-operator compression restores an equivalent structure.

The reach of the construction is set by the operator structure, not by the physics of scattering. It applies when the parameter dependence of the discretized linear system admits a low-rank, low-dimensional operator representation: the solver exports operator snapshots, SVD compresses them, and the reduced system is rebuilt differentiably. In the present implementation that condition is met by the additive depth-geometry structure of Eq.~(\ref{eq:decomp}); a comparably low-dimensional nonlinear sector would serve equally, whereas a model in which all parameters enter one coupled nonlinear function would need a different coefficient representation, and low-rank matrix snapshots alone would not remove the sampling problem. Within reaction theory the condition covers the standard coupled-channels family, including transfer and inelastic couplings, and the same logic extends to bound-state and $R$-matrix solvers whose operators depend on potential parameters. The calculation reported here is a proof of concept for the additive operator structure that phenomenological nuclear interactions typically have, not a claim of universal scaling.

The sampler comparison of Sec.~\ref{sec:cost} also shows what the gradient does and does not buy at the present size. Within the physical basin NUTS leads the ensemble sampler by about a factor of 2.4 in effective samples per model call and is close to parity once each gradient is billed at 2.2 forward evaluations, so for thirteen coordinates the gradient gives no clear efficiency gain. Thirteen is a modest dimension, and within the basin the posterior is unimodal; both samplers also accommodate linear anisotropy, NUTS through its adapted dense mass matrix and the ensemble sampler through its affine invariance, which may contribute to the small difference. Dimension is where a larger gain is expected. For product-form targets with optimally tuned step sizes, the number of leapfrog steps HMC needs per independent sample grows roughly as $d^{1/4}$ with the dimension $d$, against roughly $d$ proposals for random-walk Metropolis~\cite{Neal2011,Beskos2013}, and the stretch move of the ensemble sampler converges increasingly slowly as the dimension grows~\cite{Huijser2022}. These asymptotic results do not fix the crossover for NUTS against the ensemble sampler on a real posterior, but they place a multi-energy calibration with energy-dependent absorption, or a global potential with tens of parameters~\cite{Pruitt2023}, in the regime where the gradient should pay; a matched comparison at that size is needed to measure it. Strongly curved geometry limits HMC with a fixed mass matrix as well, so curvature alone does not guarantee a gain. The gradient is not a remedy for separated posterior structures: neither sampler mixed between those encountered here (Sec.~\ref{sec:cost}). Where the posterior is low-dimensional and unimodal, a gradient-free sampler on the same emulator can do as well. Uses of the gradient outside sampling, such as locating the posterior mode and evaluating the local curvature of the posterior, do not depend on this comparison.

Three limitations bound the physics conclusions. First, the calibration uses one angular distribution at one energy, so it determines the folded proton-plus-neutron combinations, most sharply the two total volume integrals, and cannot separate their proton and neutron contributions; the strong anti-correlations of Fig.~\ref{fig:information} are the quantitative statement of that limit. Additional energies or observables with different proton-neutron sensitivities would test whether the $15\%$ real-integral deficit persists and would put the decomposition of each total on a firmer footing; such an extension requires energy-dependent bases and a new verification of the combined reduced model, but no change to the online engine. Second, the error model is diagonal, and although Sec.~\ref{sec:results} shows the headline integrals are stable under a factor-of-two rescaling of the assigned errors, correlated normalization or model-space errors are not represented. Third, the reaction model is spinless and its continuum discretization is fixed; the posterior is conditioned on both choices, and the inferred $\beta$ absorbs their residual inadequacy only to the extent that it mimics an independent fractional error.

A joint fit of deuteron elastic scattering at several energies, with the breakup continuum coupled at each, is where the construction should pay off next. Direct gradient-free sampling of such a calibration would be prohibitively expensive, because the parameter count grows with the energy-dependent absorption while the per-solve cost stays at the coupled-channels scale; with one posterior-tested offline database per energy, the present construction moves the dominant cost to the offline stage and leaves a millisecond-scale online posterior. The same holds for propagating a calibrated deuteron potential, with its full posterior, into the transfer and surrogate-reaction analyses where deuteron optical-potential uncertainty is known to matter~\cite{Lovell2018}, and for coupling the present construction to global optical-potential frameworks with model discrepancy~\cite{Beyer2026}.

\section{Summary}
\label{sec:summary}
I have presented a strategy that makes an expensive coupled-channels reaction model usable inside gradient-based Bayesian inference without differentiating or rewriting the production solver. The solver exports parameter-dependent operator matrices and solution snapshots offline; singular value decomposition compresses them; and a reduced Galerkin model rebuilt from differentiable operations returns each likelihood value together with the exact gradient of the emulated likelihood, at a total cost of about $2.2$ forward evaluations per gradient. A verification loop that alternates inference with direct solves tests the reduced model at the posterior it produces, where its accuracy governs the reported result; it cannot detect a region supported by the direct solver that the reduced model never reveals. At thirteen sampled coordinates the gradient buys only a modest efficiency gain: within the physical basin NUTS leads a gradient-free ensemble sampler on the same emulator by about a factor of 2.4 in effective samples per model call, and the two are close to parity once the cost of the gradient is included. A clear advantage is expected for calibrations with many more parameters, such as multi-energy fits with energy-dependent absorption, where a matched comparison remains to be made; for a low-dimensional unimodal posterior a gradient-free sampler on the same emulator can do as well, and neither sampler mixed between the separated posterior structures encountered here.

Applied to $d$+$^{58}$Ni elastic scattering at $20$~MeV in a 151-channel CDCC space, the method yields a converged eighteen-input posterior from four NUTS chains in under an hour on one graphics processor, with the complete offline investment equivalent to roughly one thousand direct solves. Within the stated reaction model, continuum discretization, prior, and error assignment, the single angular distribution determines the summed real and imaginary volume integrals, $J_R/A_T=772\pm13$ and $J_W/A_T=204\pm14$~MeV\,fm$^3$; the real integral sits $15\%$ below the summed KD02 reference while the imaginary one is consistent with it, and both statements survive a global factor-of-two rescaling of the assigned experimental errors in either direction. An independent ensemble-sampler run on the same reduced model reproduces the posterior within the physical basin to one percent; the remote surrogate structure that both sampler classes encounter lies 15.5 to 32.5 log-posterior units below the physical basin at every one of twenty-five direct-solver evaluations, and the reported posterior is conditioned on the basin supported by the direct-solver evaluations. The proton and neutron contributions to each sum are strongly anti-correlated and are not separated by this observable. Multi-energy data, which the online engine accommodates without modification, are the natural next step toward resolving that decomposition.

\appendix
\section{Posterior of the sampled coordinates}
\label{app:corner}
\begin{figure*}[p]
\centering
\includegraphics[width=0.92\textwidth]{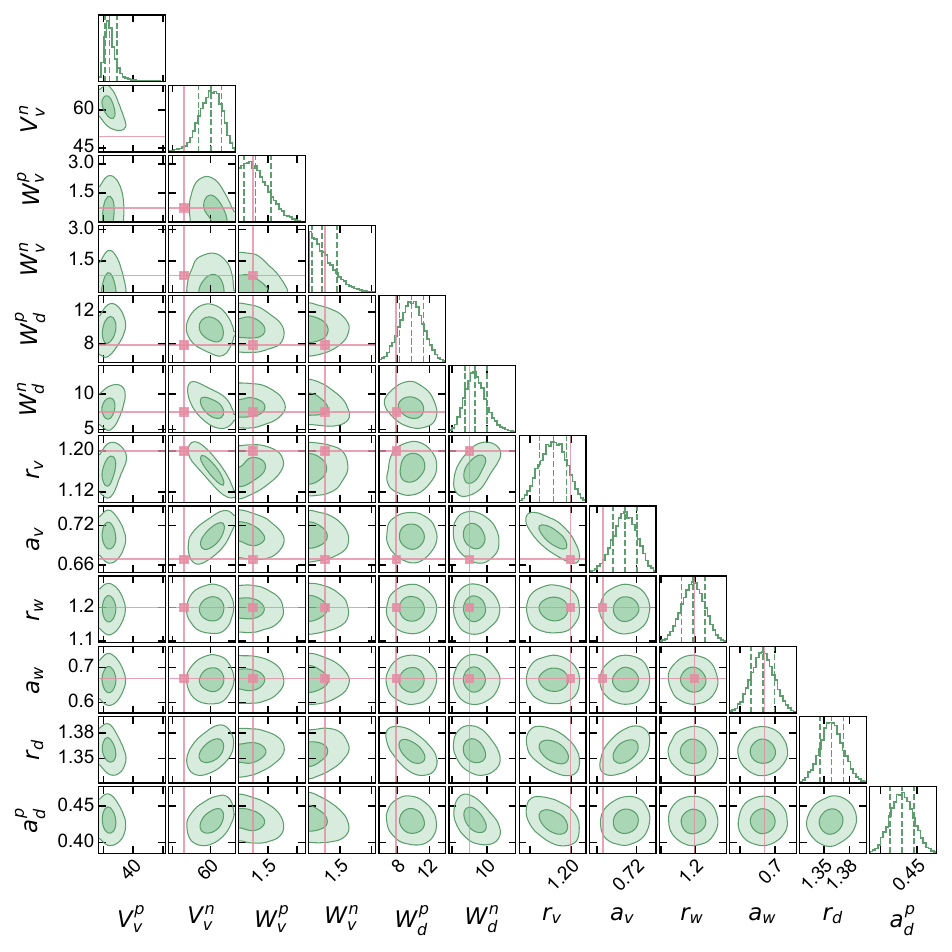}
\caption{Posterior of the twelve sampled optical coordinates for $d$+$^{58}$Ni elastic scattering at $E_\mathrm{lab}=20$~MeV. Geometry coordinates are shown by their proton members ($r_v=r_v^p$, and so on); the neutron members follow from Eq.~(\ref{eq:quantile}). Diagonal panels show marginal distributions with dashed lines at the $16\%$, $50\%$, and $84\%$ quantiles; off-diagonal panels show pairwise $1\sigma$ and $2\sigma$ contours in green. Rose lines and markers indicate the KD02 central-potential reference at $E_N=10$~MeV. Depths are in MeV; reduced radii and diffusenesses are in fm.}
\label{fig:corner}
\end{figure*}

Figure~\ref{fig:corner} shows the marginal and pairwise posterior of the twelve sampled optical coordinates: the six depths, with proton and neutron members side by side, and the six shared geometry coordinates. Through the shared-quantile map of Eq.~(\ref{eq:quantile}) each neutron geometry parameter is a fixed monotone function of its proton partner, so the remaining six of the eighteen physical inputs add no independent degree of freedom and are not shown. The depth correlations, including the proton-neutron compensation of the real depths, are posterior results.

\section{Remote structures and ensemble-sampler diagnostics}
\label{app:remote}
Two exploratory runs inform Secs.~\ref{sec:verify} and \ref{sec:cost}. A four-chain NUTS run initialized from the full prior placed three chains near $V_v^p=34$~MeV and one near $69$~MeV; the combined run failed between-chain convergence with maximum split-$\hat R=7.35$, and no chain crossed between the two structures. It served as an initialization-sensitivity test and as the cold-start comparator for the ensemble run, and ten equilibrated draws from its high-$V_v^p$ chain are among the twenty-four further points, beyond the median-nearest point of Table~\ref{tab:gate}, evaluated with the direct solver in Sec.~\ref{sec:verify}.

The ensemble run of Sec.~\ref{sec:cost} used 64 walkers for $10^5$ steps. Eighteen walkers started in the high-$V_v^p$ basin of attraction and remained there for the entire run; fourteen samples from these walkers are the remaining direct-solver points. The pooled post-burn-in diagnostics are unremarkable: integrated autocorrelation times of 500 to 910 steps, nominal effective sample sizes of 5600 to 10300 per coordinate, and a mean acceptance fraction of 0.19. Pooled over all walkers, the $V_v^p$ marginal is displaced by four standard deviations from the physical basin with six times its spread, and $J_R/A_T$ smears to $882\pm172$~MeV\,fm$^3$. Per-walker inspection reveals the trapping immediately, as would adapted ensemble diagnostics of the kind advocated in Ref.~\cite{Huijser2022}. The 46 untrapped walkers give $V_v^p=34.2\pm2.4$~MeV against $34.2\pm2.5$~MeV from the production NUTS run.

\begin{acknowledgments}
This work was supported by the National Natural Science Foundation of China (Grant Nos.~12475132 and 12535009) and the Fundamental Research Funds for the Central Universities. All numerical results were produced by the solvers and analysis scripts described in Secs.~\ref{sec:cdcc}--\ref{sec:cost}; large language models were used for language editing and code review, and the author verified all resulting text and code.
\end{acknowledgments}

\section*{Data availability}
No new experimental data were generated in this work. The measured angular distribution is publicly available from the EXFOR database (entry F0709.024) at \url{https://www-nds.iaea.org/exfor/}. The DREAM framework, with its adapter interface, offline compression, differentiable emulator, verification and rank-sweep tools, and a worked example built on a public R-matrix solver, is openly available at \url{https://github.com/jinleiphys/DREAM}. The CDCC inputs, posterior samples, and figure-generation scripts of this work are available from the author upon reasonable request.

\bibliography{references_v3}

\end{document}